\documentclass[proof]{WileyASNA-v1}

\usepackage[utf8]{inputenc}

\articletype{Article Type}%

\received{1 October 2020}
\revised{12 October 2020}
\accepted{25 October 2020}

\begin{document}

\title{The blazar sequence revised}

\author[1]{Jesús M. Rueda-Becerril*}

\author[1]{Amanda O. Harrison}

\author[1]{Dimitrios Giannios}

\authormark{\textsc{Rueda-Becerril, Harrison \& Giannios}}

\address[1]{\orgdiv{Department of Physics \& Astronomy}, \orgname{Purdue University}, \orgaddress{West Lafayette, \state{IN}, \country{USA}}}

\corres{*Jesús M. Rueda-Becerril, Department of Physics, Purdue University, 525 Northwestern Avenue, West Lafayette, IN, 47907, USA. \email{jruedabe@purdue.edu}}

\abstract{We propose and test a fairly simple idea that could account for the blazar sequence: all jets are launched with similar energy per baryon, independently of their power. For instance, flat-spectrum radio quasars (FSRQs), the most powerful jets, manage to accelerate to high bulk Lorentz factor, as observed in the radio. As a result, the emission region will have a rather modest magnetization which will induce a steep particle spectra therein and a rather steep emission spectra in the gamma- rays; particularly in the \textit{Fermi}-LAT band. For the weaker jets, namely BL Lacertae objects (BL Lacs), the opposite holds true; i.e., the jet does not achieve a very high bulk Lorentz factor, leading to more magnetic energy available for non-thermal particle acceleration and harder emission spectra. Moreover, this model requires but a handful of parameters. By means of numerical simulations we have accomplished to reproduce the spectral energy distributions and light- curves from fiducial sources following the aforementioned model. With the a complete evolution of the broadband spectra we were able to study in detail the spectral features at any particular frequency band at any given stage. Finally numerical results are compared and contrasted with observations.}


\maketitle

\footnotetext{\textbf{Abbreviations:} AGN, active galactic nucleus; EED, electron energy distribution; SED, spectral energy distribution; FSRQ, flat-specturm radio quasar; BL Lac, BL Lacertae object; PIC, particle-in-cell; BLR, broad line region}

%-------------------------------------------------------------------------------
%   INTRODUCTION
%-------------------------------------------------------------------------------
\section{Introduction}\label{sec:intro}

Radio-loud active galactic nuclei (AGNs) with a relativistic jet propagating close to the line of sight of the observer are better known as blazars. These objects have been observed in all frequencies of the electromagnetic spectrum, featuring a double bump  structure in their spectral energy distributions (SEDs). The low frequency bump peaks from infrared to X-ray, whereas the high frequency one peaks in the $\gamma$-ray. Blazars have been classified into FSRQs and BL Lacs \citep{Urry:1995pa}. Dedicated monitoring programs at different wavelengths have helped over the last decades to better understand this objects \citep[e.g.,][]{Ghisellini:2010ta,Blinov:2015pa,Lister:2016mo,Jorstad:2016ma,Ackermann2011,Rani:2017st}. For instance, thanks to \textit{Fermi}-LAT, it has been observed that BL Lacs are characterized, on average, by harder spectra than FSRQs \citep{Ghisellini:2009ma}. As a consequence, BL Lacs turn out to be the most extreme TeV emitters \citep{Ajello:2014ro}. Moreover, the synchrotron peak of BL Lacs typically characterized by a synchrotron peak at higher energies (as high as X-rays). Not surprisingly, modeling of the spectrum of blazars requires electrons injected with much higher energies in BL Lacs than in FSRQs \citep{Celotti:2008gi}. Over the last decades, radio observations have shown that FSRQs have apparent speeds $\beta_{\rm app} \sim$ tens, while for BL Lacs $\beta_{\rm app} \sim$ 2 \citep{Jorstad:2001ma,Homan:2009ka,Jorstad:2017ma,Lister:2019ho}. Also, FSRQs are likely associated with powerful jets (FR-II equivalent) in contrast to BL Lacs (FR II equivalent) \citep{Ghisellini:2001ce}. Furthermore, the luminosity of the broad-line region (BLR) may be a distinctive between blazars \citep[e.g.,][]{Ghisellini:2001ce}.

The so called \textit{blazar sequence} \citep[][]{Padovani:2007ap} has been of strong observational and theoretical focus since the first multiwavelenght SEDs of different objects were compared \citep{Fossati:1998ma}. Evolutionary scenarios have been proposed in the past decades which connect both kinds of objects in terms of accretion efficiency and the jet formation \citep{Bottcher:2002de,Maraschi:2003ta,Celotti:2008gi,Ghisellini:2011ta}.

AGN jets may develop along its way kink instabilities. This could translate into a tangled magnetic field in the jet \citep{BarniolDuran:2017tc}. Kink instabilities may induce the formation of current sheets, allowing to trigger magnetic reconnection. Recent first-principle particle in cell (PIC) simulations have demonstrated that magnetic reconnection can account for many of the extreme spectral and temporal properties of blazars \citep{Sironi:2015pe,Petropoulou:2016gi}. Interestingly, these simulations have shown that the crucial parameter that controls the distribution of accelerated particles is the jet magnetization $\sigma$. Even for a modest increase in $\sigma$ of the plasma, magnetic reconnection results in much harder particle distributions, and, as a result, harder emission spectra \citep{Petropoulou:2016gi,Petropoulou:2019si}.

The aforementioned results: MHD simulations of relativistic jets, PIC simulations of relativistic magnetic reconnection, and the kinetic features of blazar jets from radio observations; can give us information regarding the flux of matter in blazar jets. In this work we present a model and simulations that show that the baryon loading of blazar jets is similar. In Sec.~\ref{sec:model} we present our model and parametrization, and a brief description of the numerical code employed. In Sec.~\ref{sec:results} we present and describe the results obtained out of our simulations. Finally, in Sec.~\ref{sec:conclusions} we make the final conclusions from this study.

%-------------------------------------------------------------------------------
%   MODEL
%-------------------------------------------------------------------------------
\section{Model}\label{sec:model}

According to MHD theory of relativistic jets, a quantity which is conserved along magnetic field lines is the total energy flux per unit rest-mass energy flux $\mu$ \citep[see][]{Komissarov:2007ba,Tchekhovskoy:2009mc}, also known as the baryon loading. For a cold plasma flow:
\begin{equation}\label{eq:mu}
  \mu = \Gamma (1 + \sigma),
\end{equation}
where $\Gamma$ and $\sigma$ are the flow bulk Lorentz factor and magnetization, respectively. The magnetization $\sigma$ is defined as the ratio between the Poynting flux and the hydrodynamic energy flux.
\begin{equation}
  \sigma = \dfrac{B'^{2}}{4 \pi \rho' c^2},
\end{equation}
where $B'$ and $\rho'$ are the magnetic field strength and the mass density of the plasma\footnote{Quantities measured in the comoving frame of the fluid will be denoted with a prime sign ('), unless noted otherwise. Quantities measured by a cosmologically distant observer will be denoted with the subscript `obs'. Quantities measured in the laboratory frame will remain unprimed.}.

The radiative efficiency of the disk is defined as
\begin{equation}
  \eta_{\rm d} \equiv L_{\rm d} / \dot{M} c^{2},
\end{equation}
where $c$ is the speed of light, and $L_{\rm d}$ the disc luminosity. From this parameter let us define the Eddington mass accretion rate as follows:
\begin{equation}\label{eq:dotM-Edd}
  \dot{M}_{\rm Edd} \equiv \dfrac{L_{\rm Edd}}{\eta_{\rm d} c^{2}},
\end{equation}
where $L_{\rm Edd} \approx 1.26 \times 10^{36} (M / M_{\odot})$~erg~s$^{-1}$.

A main parameter in relativistic jet models is the accretion rate $\dot{M}$ onto the black hole. Let us introduce here the Eddington rate
\begin{equation}\label{eq:dotm}
  \dot{m} \equiv \dfrac{\dot{M}}{\dot{M}_{\rm Edd}},
\end{equation}
where $\dot{M}_{\rm Edd}$ is the Eddington mass accretion rate. Therefore, $\dot{m}$ gives a measure of the accretion rate of the AGN as a fraction of the Eddington rate. The jet luminosity $L_{\rm j}$ is related to the accretion power by
\begin{equation}\label{eq:Lj-dotM}
  L_{\rm j} = \eta_{\rm j} \dot{M} c^{2}
\end{equation}
where $\eta_{\rm j}$ is the jet production efficiency. From this and \eqref{eq:dotM-Edd} we get that
\begin{equation}\label{eq:Mdot-Medd-rat}
  L_{\rm j} = \dfrac{\eta_{\rm j}}{\eta_{\rm d}} L_{\rm Edd} \dot{m}.
\end{equation}

According to radio observations there seems to be a correlation between the bulk Lorentz factor of the emission region and the jet power \citep{Lister:2009co,Homan:2009ka}. According to Eq.~\eqref{eq:Mdot-Medd-rat}. Out of these empirical relation we make the following ansatz:
\begin{equation}\label{eq:dotm-gamma}
  \dot{m} = {\left( \dfrac{\Gamma}{\Gamma_{0}} \right)}^{s}.
\end{equation}
From empirical results we have set $s \sim 3$ and $\Gamma_{0} \sim 40$. For further details on this regard see \citet{RuedaBecerril:2020ha}.

The Poynting flux luminosity of the jet is given by
\begin{equation}\label{eq:LB-sigma-Lj}
  L_{\rm B} = \dfrac{\sigma}{1 + \sigma} L_{\rm j},
\end{equation}
which in turn we use to calculate the magnetic energy density of the emitting blob in the comoving frame:
\begin{equation}\label{eq:uB}
  u'_{\rm B} = \dfrac{L_{\rm B}}{2 \pi R_{\rm b}'^2 c \beta \Gamma^{2}},
\end{equation}
where $R_{\rm b}'$ is the size of the emission region. From this results we get that the luminosity of the electrons in the comoving frame of the emitting plasma reads:
\begin{equation}\label{eq:Le1}
  L_{\rm e}' = f_{\rm rec} \dfrac{2 L_{\rm B}}{3 \beta \Gamma^{2}}
\end{equation}

Regarding the broad line region (BLR), in our model we will assume that the emission region is immersed in an isotropic and monochromatic radiation field. The energy density of the external BLR radiation can be parametrized as follows \citep{Ghisellini:2008ta}:
\begin{equation}\label{eq:uBLR}
  u_{\rm BLR} = \eta_{\rm BLR} \dfrac{L_{\rm d}}{4 \pi c R_{\rm BLR}^2}
\end{equation}
where $R_{\rm BLR} \simeq 10^{17} L_{{\rm d},45}^{1 / 2}$~cm is the radius of the BLR, $\eta_{\rm BLR}$ the covering factor, and $L_{{\rm d}, 45} = L_{\rm d} / (10^{45} \, \mathrm{erg \, s^{-1}})$. Finally, we will consider the radiation field in this region to be monochromatic with frequency $\nu_{\rm BLR}$. In the comoving frame of the plasma flow, $\nu'_{\rm BLR} = \nu_{\rm BLR} \Gamma$ and $u'_{\rm BLR} = u_{\rm BLR} \Gamma^2 (1 + \beta^2 / 3)$, where $\beta \equiv \sqrt{1 - \Gamma^{-2}}$ is the bulk speed of the flow in units of the speed of light.

In the emission region we assume that magnetic reconnection takes place. A fraction of this energy $f_{\rm rec}$ goes into accelerated electrons due to this process. From the average energy and average Lorentz factor of the injected electrons one finds that
\begin{equation}\label{eq:gmin-frec}
  \gamma_{\min}' = f_{\rm rec} \sigma \dfrac{m_{\rm p}}{m_{\rm e}} \left(\dfrac{p - 2}{p - 1}\right).
\end{equation}
The above result holds for $p > 2$ and $\gamma_{\max}' \gg \gamma_{\min}'$. On the other hand, if the distribution has a power-law index of $1 < p < 2$ we can make use of \citep{Sironi:2014sp}
\begin{equation}\label{eq:gmax-sigma}
  \gamma_{\max}' = {\left(f_{\rm rec}(\sigma + 1) \dfrac{m_{\rm p}}{m_{\rm e}} \dfrac{2 - p}{p - 1}\right)}^{1 / (2 - p)}{\gamma_{\min}'}^\frac{1 - p}{2 - p}.
\end{equation}

Finally, regarding the value of $\gamma_{\max}'$ for the cases with $p > 2$ is estimated by equating the acceleration rate of the electrons to the synchrotron cooling rate \citep{Dermer:2009}, i.e.,
\begin{equation}\label{eq:gmax-eacc}
  \gamma_{\max}' = {\left( \dfrac{6 \pi {\rm e}}{ \epsilon_{\rm acc} \sigma_{\rm T} B'} \right)}^{1/2},
\end{equation}
where the parameter $\epsilon_{\rm acc}$ could be interpreted as the number of gyrations the electron experience before it is injected into the system as part of the non-thermal distribution.

We perform our simulations using the numerical code \texttt{Paramo} \citep{RuedaBecerril:2020kn}. This code solves the Fokker-Planck equation using a robust implicit method \citep[see][]{Chang:1970co,Park:1996pe}, and for each time-step of the simulation the synchrotron, synchrotron self-absorption and inverse Compton emission (both synchrotron self-Compton, SSC, and external Compton, EIC) are computed with sophisticated numerical techniques \citep{Mimica:2012al,RuedaBecerril:2017mi}. For the present work we will focus on solving the Fokker-Planck equation without diffusion terms, i.e.,
\begin{equation}\label{eq:FP}
  \dfrac{\partial n'(\gamma', t')}{\partial t'} + \dfrac{\partial}{\partial\gamma'} \left[ \dot{\gamma}'(\gamma', t') n'(\gamma', t') \right] = Q(\gamma', t') - \dfrac{n'(\gamma', t')}{t_{\rm esc}},
\end{equation}
where $n'$ is the electrons energy distribution (EED) in the flow comoving frame, $Q$ is a source term, and $t_{\rm esc}$ is the electrons the average escape time. The electrons radiative energy losses are accounted for with the coefficient:
\begin{equation}\label{eq:ener-loss}
  -\dot{\gamma}' = \dfrac{4 c \sigma_{\rm T}}{3 m_{\rm e} c^2} \beta_{\rm e}'^2 \gamma'^2 (u_{\rm B}' + u_{\rm BLR}'),
\end{equation}
where $\beta'_{\rm e}$ is the speed of the electron, in units of $c$, in the comoving frame.
%-------------------------------------------------------------------------------

%-------------------------------------------------------------------------------
%   RESULTS
%-------------------------------------------------------------------------------
\section{Results}\label{sec:results}

\begin{table}
  \centering
  \begin{tabular}{cc}
    Parameter & Value \\
    \hline
    \hline
    $\theta_{\rm obs}$ & $2^{\circ}$ \\
    $M_{\rm bh}$       & $10^{9} M_{\odot}$ \\
    $\eta_{\rm j}$     & 0.9 \\
    $\eta_{\rm d}$     & 0.1 \\
    $\eta_{\rm BLR}$   & 0.1 \\
    $\nu_{\rm BLR}$    & 2 eV / $h$ \\
    $f_{\rm rec}$      & 0.15 \\
    $s$                & 3.0 \\
    $\Gamma_0$         & 40 \\
    $\mu$              & 50, 70, 90 \\
    ($\sigma, p$)      & (1, 3.0), (3, 2.5), (10, 2.2), (15, 1.5), (20, 1.2)
  \end{tabular}
  \caption{Parameters of the present model. See text for a description of each of them.}
  \label{tab:tab1}
\end{table}

\begin{figure*}
  \centering
  \includegraphics[width=0.98\textwidth]{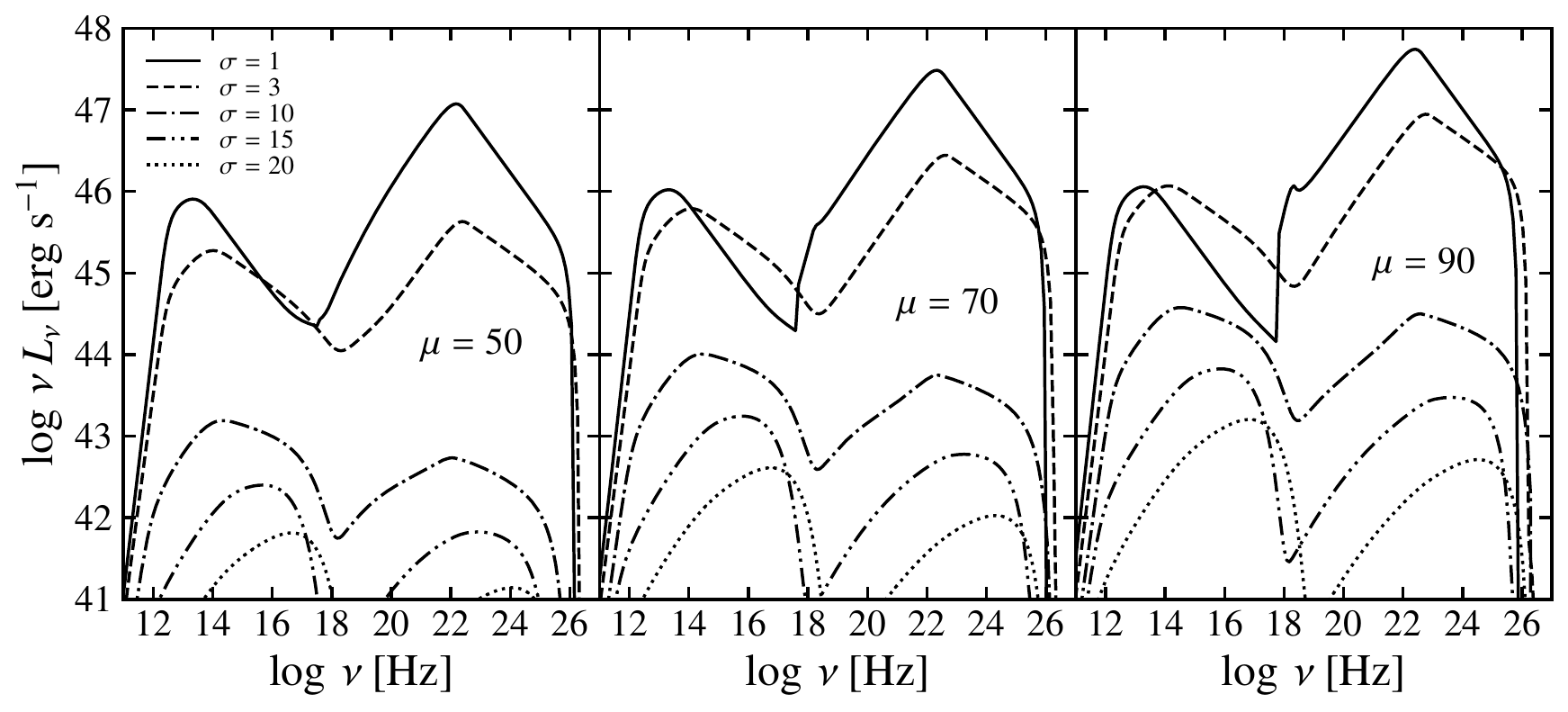}
  \caption{Sequence of blazar SEDs for varying model parameters. From left to right, each panel shows the averaged SEDs for different baryon loading $\mu = 50, 70, 90$, respectively. Solid, dashed, dot-dashed, dot-dot-dashed and dotted lines correspond to those simulations with $\sigma = 1, 3, 10, 15, 20$, respectively. The SEDs were averaged over 1~day after particles start being injected in the emission region.}
  \label{fig:BlazSeq}
\end{figure*}

In Tab.~\ref{tab:tab1} we summarize the parameters and values employed in the present work.
\begin{figure*}
  \centering
  \includegraphics[width=0.98\textwidth]{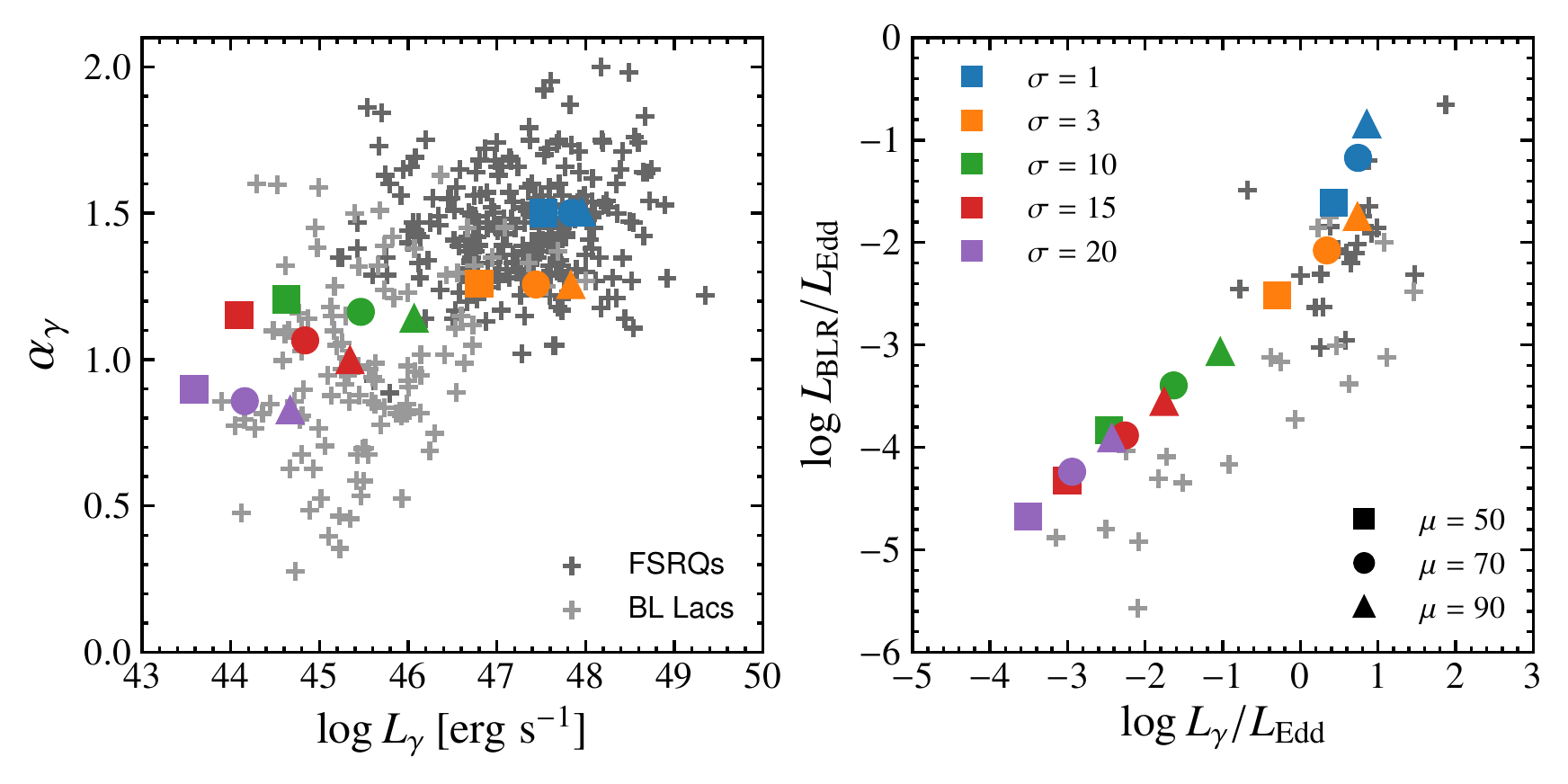}
  \caption{{\it Left panel}: $\gamma$-ray energy spectral index $\alpha_{\gamma}$ as a function of the $\gamma$-ray luminosity $L_{\gamma}$. {\it Right panel}: Luminosity of the BLR $L_{\rm BLR}$ as a function of $L_{\gamma}$, both in units of the Eddington luminosity $L_{\rm Edd}$.}
  \label{fig:specidx}
\end{figure*}
Our model resides on the hypothesis that all blazars are launched with similar baryon loading. In Fig.~\ref{fig:BlazSeq} we show the sequence of SEDs for three different values of $\mu$. The solid, dashed, dot-dashed, dot-dot-dashed and dotted lines correspond to magnetization $\sigma = 1, 3, 10, 15, 20$, respectively. FSRQs are the brightest of all blazars in all frequencies, their inverse Compton (IC) component tends to be louder than the synchrotron one, and $\nu_{\rm syn}$ falls in the infra-red. These features also appear in our simulations with the lowest magnetization, which we assumed as FSRQ-like. On the other hand, the main features observed in SEDs of BL Lac objects are a quieter IC component, $\nu_{\rm syn}$ in the UV--X-rays, and a harder spectral index in the $\gamma$-rays. We find that this is also the case for the highly magnetized cases. Finally, by contrasting all frames in Fig.~\ref{fig:BlazSeq} we can see the \textit{blazar sequence} trend \citep[cf. ][Fig.~12]{Fossati:1998ma} is favored for $\mu > 50$. The jets with larger baryon loading correspond to those sources with larger bulk Lorentz factor. From Eq.~\eqref{eq:dotm-gamma}, these sources correspond to the most efficient accretion disks which in turn correspond to those with most powerful jets (see Eq.~\eqref{eq:Mdot-Medd-rat}). This effect is more evident for the highly magnetized cases, whose luminosity increases for almost two orders of matnitude.

In Fig.~\ref{fig:specidx} we present our simulations with $\sigma = 1, 3, 10, 15, 20$ in blue, orange, green, red and purple points, respectively. Those simulations with baryon loading $\mu = 50, 70, 90$ are depicted in squares, circles and triangles, respectively. Observation data from \cite{Ghisellini:2011ta} is seen in light and dark gray crosses. On the left panel, we show the spectral index $\alpha_{\gamma}$ as a function of the bolometric luminosity $L_{\gamma}$ in the band 0.1-10~GeV \citep[cf. Fig.~1 in][]{Ghisellini:2011ta}.

On the right panel of Fig.~\ref{fig:specidx} we show the BLR luminosity, $L_{\rm BLR}$, as a function of $L_{\gamma}$, both in units of the Eddington luminosity $L_{\rm Edd}$, together with observational data points from  Fig.~3 in \citet[][]{Ghisellini:2011ta}. According to \citet{Ghisellini:2011ta}, those sources with a stronger emission lines, i.e., showing a more luminous BLR, appear louder in the $\gamma$-ray band. The latter being FSQRs. In our simulations, the corresponding ones with a more luminous BLR are those with larger $\Gamma$. Our model states that these objects have larger Eddington ratio (see Eq.~\eqref{eq:dotm-gamma}), i.e., that would correspond to highly efficient accretion objects.

\begin{figure*}
  \centering
  \includegraphics[width=0.98\textwidth]{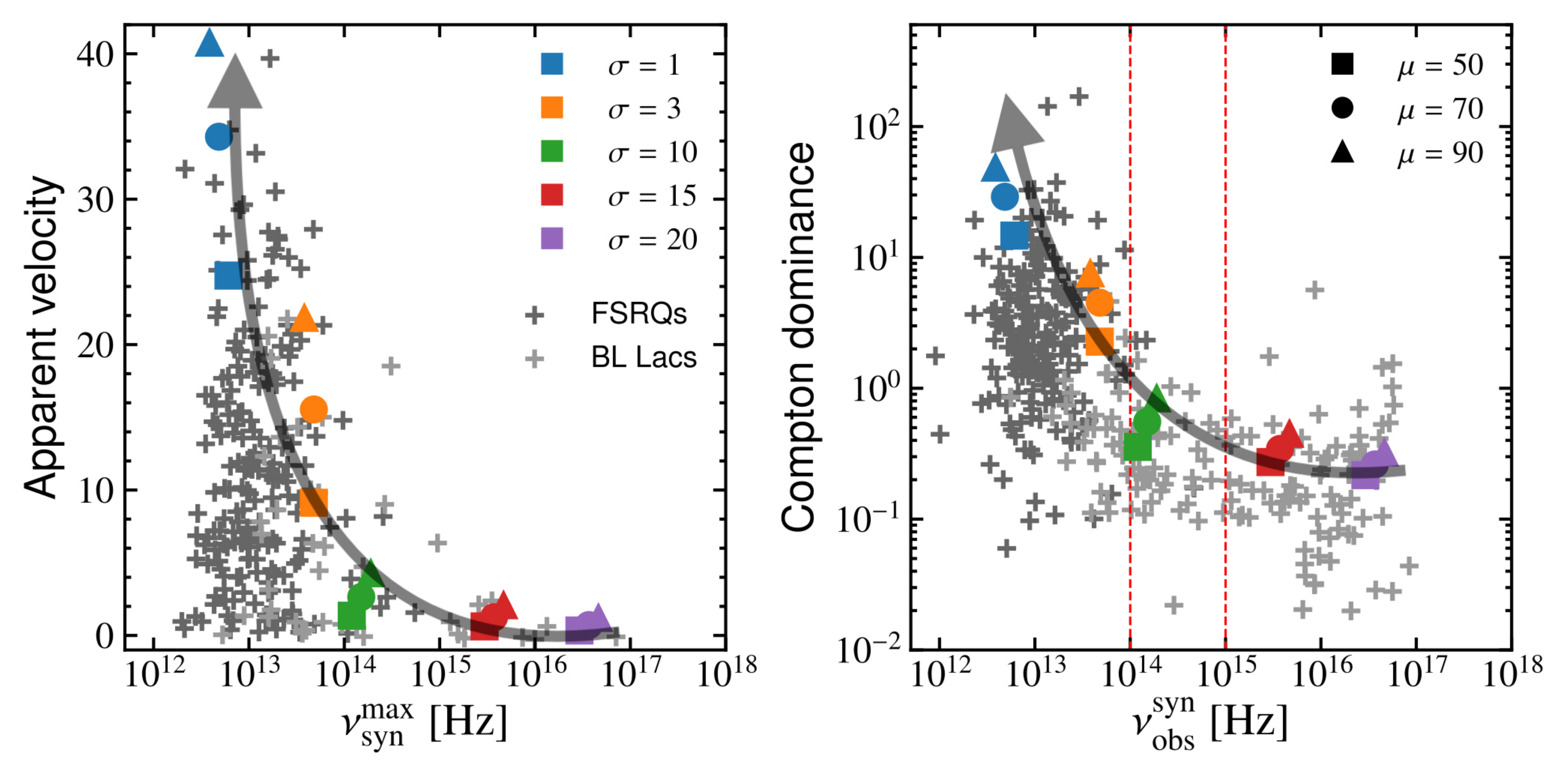}
  \caption{{\it Left panel}: We show the apparent velocity as a function of the synchrotron peak frequency $\nu_{\rm syn}$. Observational data from \citet{Lister:2019ho}. {\it Right panel}: We show the Compton dominance as a function of the synchrotron peak. In red dashed vertical lines we separate the LBL ($\lesssim 10^14$~Hz), IBL ($\gtrsim 10^{14}~Hz$ and $\lesssim 10^{15}$~Hz) and HBL ($\gtrsim 10^{15}$~Hz) regions. Observational data from \citet{Finke:2013co}.}
  \label{fig:G-vsyn}
\end{figure*}

In the same manner, in Fig.~\ref{fig:G-vsyn} we present our simulations with $\sigma = 1, 3, 10, 15, 20$ in blue, orange, green, red and purple points, respectively. Baryon loadings $\mu = 50, 70, 90$ are shown in squares, circles and triangles, respectively. Light and dark gray crosses correspond to BL Lacs and FSRQs sources, respectively. On the left panel we show the apparent velocity of our synthetic objects. The observational data correspond to the data in the MOJAVE survey, reported in \citet{Lister:2019ho}. A translucent gray arrow draws the trend of increment of the jet luminosity. In this plot we can appreciate how the synchrotron peak $\nu_{\rm syn}$ of our simulations is similar for each magnetization. The apparent velocity is bulk Lorentz factor dependent due to relativistic boosting. This effect is clear for those objects with larger $\Gamma$ (blue and orange points), which correspond to those simulations with more powerful jets. Our simulations with powerful jets concur with FSRQs as assumed. This is the case as well with highly magnetized objects. These objects represent the less powerful jets, and fall well in the region of BL Lacs.

The \textit{Compton dominance} is defined as the ratio of luminosities between the IC and the synchrotron components of their SED. On the right panel of Fig.~\ref{fig:G-vsyn} we contrast the Compton dominance and $\nu_{\rm syn}$ of our synthetic sources with the observational data reported in \citet{Finke:2013co}, depicted as gray crosses. These sources are presented in the 2LAC clean sample where all had known redshift and could clearly be classified. In that same work, sources with unknown redshift were also taken into account, finding that the relation between Compton dominance and synchrotron peak frequency have a physical origin rather than it being a redshift selection effect. Regarding our simulations, we can observe that all our simulations fall within the observational points. The gray transparent arrow shows the trend of increment of the jet luminosity. Our simulations show that, keeping $\mu$ constant, changing the magnetization will give the transition from synchrotron-dominant (highly magnetized) to Compton-dominant and $\gamma$-ray loud sources.
%-------------------------------------------------------------------------------

%-------------------------------------------------------------------------------
%   CONCLUSIONS
%-------------------------------------------------------------------------------
\section{Conclusions}
\label{sec:conclusions}

Our model assumes that all jets are injected with energy per baryon in a narrow range $50 \lesssim \mu \lesssim 80$ and that the jet bulk Lorentz factor and power scale positively with the accretion rate, and can account for or predict:
\begin{itemize}
  \item That $\dot{m}$ controls many of the observable features of blazars such as the high-energy spectral index and luminosity, the brightness of the BLR, the apparent speed, and the synchrotron spectrum and synchrotron peak frequency.
  \item Sources that are $\gamma$-ray brighter have softer $\gamma$-ray spectral index $\alpha_{\gamma}$. Lower values of $\alpha_{\gamma}$ (i.e., harder spectra) were found for the $\gamma$-ray quieter sources.
  \item The BLR luminosity $L_{\rm BLR}$ scales linearly with the $\gamma$-ray luminosity of the object.
  \item Fastest objects have low-frequency synchrotron peak $\nu_{\rm syn}$ while objects with intermediate-to-high synchrotron peak move rather slow.
  \item Low jet luminosity sources are non-Compton dominant but high synchrotron-peaked, whereas those with higher Compton dominance have a $\nu_{\rm syn} \lesssim 10^{13}$~Hz.
\end{itemize}

%----------------------------------------------------------------------------------------
%	ACKNOWLEDGEMENTS
%----------------------------------------------------------------------------------------
\section*{Acknowledgments}

The research was partly supported by \fundingAgency{NASA Fermi Cycle 12 Guest Investigator Program} \#\fundingNumber{121077}. JMRB acknowledges the support from the Mexican National Council of Science and Technology (\fundingAgency{CONACYT}) with the Postdoctoral Fellowship under the program Postdoctoral Stays Abroad \fundingNumber{CVU332030}. DG acknowledges support from the \fundingAgency{NASA ATP} \fundingNumber{NNX17AG21G}, the \fundingAgency{NSF} \fundingNumber{AST-1910451} and the \fundingAgency{NSF} \fundingNumber{AST-1816136} grants. This research was supported in part through computational resources provided by Information Technology at Purdue, West Lafayette, IN, USA.

%----------------------------------------------------------------------------------------
%	REFERENCES
%----------------------------------------------------------------------------------------
\bibliography{refs}%

\end{document}